\documentclass[11pt,notitlepage,prd,a4paper,%preprint,
preprintnumbers,superscriptaddress,nofootinbib]{revtex4-1}
\usepackage[utf8]{inputenc}
\usepackage[colorlinks=true,citecolor=blue,linkcolor=blue]{hyperref}
\usepackage[normalem]{ulem}
\usepackage{url}
\usepackage{graphicx,wrapfig,float,slashed,cancel}
\usepackage{amsmath,amssymb,epsfig,graphicx,xcolor,stmaryrd}
\usepackage{bm}
\usepackage{enumitem}
\usepackage{multirow}

\usepackage{natbib}
\usepackage{amsmath,amssymb,enumitem}
\usepackage{braket}
\usepackage{fontenc}[T1]
\usepackage[utf8]{inputenc}
\usepackage{multirow}
\usepackage{placeins}
\usepackage{slashed}
\usepackage{braket}
\usepackage{xspace}

\newcommand{\be}{\begin{equation}}
\newcommand{\ee}{\end{equation}}
\newcommand{\refeq}[1]{Eq.~(\ref{eq:#1})}

\makeatletter

\def\ak{\@ifstar\@@ak\@ak}
\newcommand{\@ak}[1]{\textcolor{ForestGreen}{[\textbf{AK:} #1]}}
\newcommand{\@@ak}[1]{\textcolor{ForestGreen}{#1}}
\makeatother

\definecolor{darkblue}{RGB}{1, 90, 173}

\begin{document}

%%%%%%%%%%%%%%%%%%%%%%%%%%% ---------- Title ---------- %%%%%%%%%%%%%%%%%%%%%%%%%%%

\title{Analysis of the $B_{(s)}\rightarrow T(J^P=2^-)$ transition in light cone QCD sum rules}

\author{T.~M.~Aliev}
\email{ taliev@metu.edu.tr}
\affiliation{Physics Department, Middle East Technical University, 06531, Ankara, T\"{u}rkiye}
\author{Y.~Sarac}
\email{ yasemin.sarac@atilim.edu.tr}
\affiliation{Electrical and Electronics Engineering Department,
Atilim University, 06836 Ankara, T\"{u}rkiye}

\date{\today}

\preprint{}

\begin{abstract}

The semileptonic $B_{(s)} \rightarrow T(J^P=2^-)l^+l^-$ decays induced by flavor changing neural currents are investigated within the light cone QCD sum rule method in the leading order of $O(\alpha_s)$. We apply the $B$ meson distribution amplitudes up to twist-4 and calculate the relevant form factors of the $B_{(s)} \rightarrow T$ transitions, where  $T=K_2,~a_2,~f_2,~\phi_2$ with  $J^P=2^-$. The obtained results of the form factors then adopted in the calculations of the corresponding widths. The present results can be used in future experiments for studying the properties of $J^P=2^-$ tensor mesons.

\end{abstract}

%%%%%%%%%%%%%%%%%%%%%%%%%%%%%%%%%%%%%%%%%%%%%%%%%%%%%%%%%%%%%%%%%%%%%%%%%%%%%%%%%%

\maketitle

%\vspace{-1mm}
%%%%%%%%%%%%%%%%%%%%%%%%%%%%%%%%%
%%%%%%%%%%%%%%%%%%%%%%%%%%%%%%%%%
%\maketitle
\renewcommand{\thefootnote}{\#\arabic{footnote}}
\setcounter{footnote}{0}
%%%%%%%%%%%
\section{\label{sec:level1}Introduction}\label{intro}

Investigations of the $B$ meson decays are among promising tools to verify the gauge structure of the Standard Model (SM), explore potential new physics beyond it, and precisely determine the elements of the Cabibbo-Kobayashi-Maskawa (CKM) matrix. It is due to the large mass of $b$ quarks where perturbative QCD works well, hence predictions are reliable. At the same time flavor-changing decays of $B$ meson are very sensitive to the existence of new physics.

The $B$ meson transition to tensor mesons contain additional kinematical quantities resulting from the additional polarization of the tensor meson compared to the vector mesons provide us with an advantage in the search of the NP in the $B$ meson to tensor transitions. In this respect, these transitions can be taken as an additional platform to identify new helicity structures that differ from the predictions of the SM.

Here we would like to make the following remark. In SM the higher dimension operators are suppressed by power of $m_W$ and usually are neglected. The higher dimension operators form from higher spin in quark and lepton sectors can bring new features. An operator with spin j can be represented by

\begin{eqnarray}
O^j =\bar{s}_L\Gamma^{(j-)}_{\mu_1\cdots \mu_j}b_L\bar{l}\Gamma^{(j+)\mu_1\cdots \mu_j}l,
\label{eq:fitform}
\end{eqnarray}    
Where $\Gamma^{(\pm)}_{\mu_1\cdots \mu_j}=\gamma_{\{\mu_1}D^{\pm}_{\mu_2}\cdots D^{\pm}_{\mu_j\}}$,  
$ D^{\pm}=\overrightarrow{D}\pm \overleftarrow{D}$ and curly bracket indicated symmetrization in the Lorentz indices. Relative contributions of this matrix elements are in order $(m_B/m_W)^{2(j-1)}$. Operators of the form give new possibility to test physics beyond the SM (see references for example \cite{Gratrex:2015hna}, \cite{Fajfer:2018lix}). In the present work contributions of these operators are neglected.

The transitions $B \rightarrow T$ were considered in various works for the final tensor meson states with $J^P=2^+$ and corresponding form factors were obtained by applying various methods~\cite{Scora:1995ty,Charles:1999gy,Safir:2001cd,Ebert:2001pc,Cheng:2003sm,Datta:2007yk,Cheng:2009ms,Cheng:2010sn,Sharma:2010yx,Wang:2010ni,Wang:2010tz,Yang:2010qd,Azizi:2013aua,Khosravi:2015jfa,Emmerich:2018rug,Bernlochner:2017jxt,Aliev:2019ojc,Zuo:2021kui,Chen:2021ywv,Zuo:2024jdf,Vardani:2024bae,Salam:2024nfv}. In Ref.~\cite{Emmerich:2018rug}, the $B\rightarrow f_2(1270)$ transition was studied via the LCSR method with the DAs of the $f_2(1270)$ meson. The LCSR framework was also applied for the $B\rightarrow f_2,~a_2,~K_2^{*},~f'_2$ transitions via the DAs of the considered tensor-mesons~\cite{Yang:2010qd}. Another similar investigation using the LCSR method was presented in Ref.~\citep{Wang:2010tz} for $B\rightarrow f_2,~a_2,~K_2^{*}$ in which only the $\phi_+,~\bar{\phi}$ DAs of the $B$ meson were used. These form factors were also calculated in Ref.~\cite{Wang:2010ni} by applying the perturbative QCD approach. In Ref.~\cite{Khosravi:2015jfa}, form factors for the $B\rightarrow f_2,~a_2,~K_2^{*}$ transitions, and in Ref.~\cite{Azizi:2013aua}, those for $B\rightarrow D_2^{*}$ transition were calculated using the three-point QCD sum rule method. In Ref.~\cite{Chen:2021ywv}, the $B$ and $B_s$ mesons transitions to tensor mesons, $B\rightarrow (a_2,~K_2^*,~D_2^*)$ and  $B_s\rightarrow (K_2^*,~f_2',~D_{s2}^*)$, were considered and their corresponding form factors were calculated using the standard and covariant light-front quark model. In Ref.~\cite{Aliev:2019ojc}, $B\rightarrow T$ transition for the tensor mesons $D_2,~f_2,~a_2$, and $~K_2^{*}$ were considered, and the corresponding form factors for the tree-level and flavor-changing neutral current transitions were obtained using the LCSRs approach. The LCSRs analyses were also performed in Refs.~\cite{Zuo:2021kui,Zuo:2024jdf} for light tensor mesons with $J^{PC}=2^{++}$ ($a_2(1320),~K_2^*(1430),~f_2(1270),~f_2'(1525)$) and $J^{PC}=2^{--}$ ($\rho_2,~\omega_2,~K_2,~\phi_2$) mesons by using distribution amplitudes of the tensor mesons.

In the present work, we aim to calculate the form factors for the $B$ meson to negative parity tensor meson ($J^{PC}=2^{--}$) transitions, which are among the main ingredients for the investigations of these transitions. To this end, we apply the light-cone QCD sum rules (LCSRs)~\cite{Balitsky:1989ry,Chernyak:1990ag} and use $B$-meson light-cone distribution amplitudes (LCDAs). For the review about the LCSR method, we refer the reader to Ref.~\cite{Colangelo:2000dp}. The recent applications of the LCSR method to the heavy baryon and heavy meson physics can be seen in Refs.~\cite{Khodjamirian:2006st,Faller:2008tr,Wang:2015vgv,Wang:2015ndk,Cheng:2017smj,Gubernari:2018wyi,Gao:2019lta,Cheng:2019tgh,Descotes-Genon:2019bud}.

In the present article, we have the following outline: In the next section, Sec.~\ref{II}, we obtain the LCSRs for the considered transitions. Section~\ref{III} presents the numerical analysis of the results for the form factors attained in Sec.~\ref{II}. Section~\ref{IV} provides a summary of the obtained results. In the Appendix, supplementary materials, namely the two-particle DAs of the $B$ meson and analytic results corresponding to the calculations are presented.  

\section{$B\rightarrow T$ transition form factors in light cone QCD sum rules }\label{II}

In this section, the light cone sum rules (LCSR) for the form factors responsible for the $B_{(s)}\rightarrow T$ transitions, where $T$ represents the tensor mesons $T=K_2,~a_2,~f_2,~\phi_2$ with $J^{P}=2^{-}$, are derived.

In order to obtain the form factors, we define the following correlation function:
\begin{eqnarray}
 \label{eq:correlator}
    \Pi^{\mu\nu\rho}(q, k)
        = i \int \text{d}^4 x\, e^{i k\cdot x}\,
        \bra{0} \mathcal{T} \{ J^{\mu\nu}(x), J^{\rho}(0)\} \ket{B_{q_2}(p)}\,
\end{eqnarray}
where $J^{\mu\nu}=\bar{q}_2\Gamma_1^{\mu\nu}q_1$ is the interpolating current for corresponding tensor meson and $J^{\rho}=\bar{q}_1\Gamma_{2}^{\rho}h_v$ is the weak current with momentum $q$, $h_v$ is the heavy quark effective field for $b$ quark and $p$ is the $B$ meson state momentum. In the correlator, we need the proper interpolating current for the considered tensor meson, which is
\begin{eqnarray}
J^{\mu\nu}=\frac{i}{2}(\bar{q}_2 \gamma^{\mu}\gamma_5\overleftrightarrow{\mathcal{D}}^{\nu}q_1+\bar{q}_2 \gamma^{\nu}\gamma_5\overleftrightarrow{\mathcal{D}}^{\mu}q_1),\label{eq:current1}
\end{eqnarray}
for $K_2$, $\phi_2$, and $a_2$ tensor mesons and
\begin{eqnarray}
J^{\mu\nu}=\frac{i}{2\sqrt{2}}(\bar{q}_2 \gamma^{\mu}\gamma_5\overleftrightarrow{\mathcal{D}}^{\nu}q_1+\bar{q}_2 \gamma^{\nu}\gamma_5\overleftrightarrow{\mathcal{D}}^{\mu}q_1+\substack{q_1\rightarrow q_1'\\q_2\rightarrow q_2'}),\label{eq:current2}
\end{eqnarray}
for $f_2$ meson with $\overleftrightarrow{\mathcal{D}}^{\nu}=\frac{1}{2}[\overrightarrow{\mathcal{D}}^{\nu}-\overleftarrow{\mathcal{D}}^{\nu}]$ and $\overrightarrow{\mathcal{D}}^{\nu}=\overrightarrow{\mathcal{\partial}}^{\nu}-ig\frac{\lambda^a}{2}A_a^{\nu}$ where $\mathcal{\partial}^{\nu}=\frac{\partial}{\partial x_{\nu}}$. The $q_1(q_1')$, $q_2(q_2')$ and $\Gamma_2^{\rho}$ in Eqs.~(\ref{eq:current1}), (\ref{eq:current2}) and the transition current  are presented in Table~\ref{tab:int.current}.
\begin{table}[]
\begin{tabular}{|c|c|c|c|c|}
\hline
Transition          & $q_1(q_1')$ & $q_2(q_2')$ & $\Gamma_2^{\rho}$                                                                                                                                          & Form factor                                                                                         \\ \hline\hline
$B\rightarrow K_2$  & $s$         & $d$         & \multirow{4}{*}{\begin{tabular}[c]{@{}c@{}}$\gamma^{\rho}\gamma_5$\\ $\gamma^{\rho}$\\ $\sigma^{\rho\{q\}}\gamma_5$\\ $\sigma^{\rho\{q\}}$\end{tabular}} & \multirow{4}{*}{\begin{tabular}[c]{@{}c@{}}$A$\\ $V_0,~V_1,~V_2$\\ $T_1$\\ $T_2,~T_3$\end{tabular}} \\ \cline{1-3}
$B_s\rightarrow \phi_2$ & $s$         & $s$         &                                                                                                                                                          &                                                                                                     \\ \cline{1-3}
$B\rightarrow f_2$      & $d(u)$      & $d(u)$      &                                                                                                                                                          &                                                                                                     \\ \cline{1-3}
$B\rightarrow a_2$      & $d$         & $u$         &                                                                                                                                                          &                                                                                                     \\ \hline
\end{tabular}
\caption{The quark fields present in the interpolating currents $J^{\mu\nu}$ and $J^{\rho}$ and $\Gamma_2^{\rho}$ present in $J^{\rho}$. }
\label{tab:int.current}
\end{table}

The correlation function can be written in terms of hadronic parameters and can also be calculated using the light cone operator product expansion (OPE). The LCSR for the form factors is obtained by matching the hadronic and OPE results for the correlation function using the quark-hadron duality ansatz. 

Let us first calculate the hadronic part of the correlation function. The hadronic dispersion relation of the correlation function is obtained by calculating the imaginary part of it with respect to the variable $k^2$ by inserting a complete set of hadrons (in our case, tensor mesons). In result, we have
\begin{equation}
\textrm{Im}\Pi^{\mu\nu\rho}=\pi\delta(s-m_T^2) \sum \bra{0} J^{\mu\nu} \ket{T(k,\varepsilon)}\bra{T(k,\varepsilon)} J^{\rho} \ket{B(p)},
\label{Eq:corr}
\end{equation}
where summation means summation over the polarization of tensor the meson. The first matrix element, $ \bra{0} J^{\mu\nu} \ket{T(k,\varepsilon)}$, is defined in term of decay constant as
\begin{eqnarray}
  \bra{0} J^{\mu\nu} \ket{T(k,\varepsilon)}=f_Tm_T^3\varepsilon_{\mu\nu},
 \label{Eq:Matrix1}
\end{eqnarray}
where $m_T$ and $\varepsilon_{\mu\nu}$ are the mass and polarization tensor of the tensor meson. The second matrix element, $\bra{T(k,\varepsilon)} J^{\rho} \ket{B(p)}$, in terms of form factors is parameterized as
\begin{eqnarray}
    \label{eq:BtoT:avector}
    \bra{T(k, \varepsilon)} \bar{q}_1 \gamma^\rho \gamma_5 b \ket{B(p)}
         =   2 \epsilon^{\rho\beta\delta\sigma} \varepsilon^*_\beta p_\delta k_\sigma \frac{A}{m_B + m_T}\,,
    \label{eq:BtoT:axial}
\end{eqnarray} 
\begin{eqnarray}
      \bra{T(k, \varepsilon)} \bar{q}_1 \gamma^\rho  b \ket{B(p)}
         = i \varepsilon^*_\beta\ \bigg[g^{\rho\beta} (m_B + m_T) V_1 - \frac{(p + k)^\rho q^\beta}{m_B + m_T}\, V_2
        - q^\rho q^\beta \frac{2 m_T}{q^2} \left( { V}_3 - { V}_0\right)\bigg]\,,
    \label{eq:BtoT:tensor}
\end{eqnarray}    
\begin{eqnarray}    
      \bra{T(k, \varepsilon)} \bar{q}_1  \sigma^{\rho\alpha}\, q_\alpha \gamma_5 b \ket{B(p)}
         = -  2  i \epsilon^{\rho\beta\delta\sigma} \varepsilon^*_\beta p_\delta k_\sigma T_1\,,
    \label{eq:BtoT:tensor5}
\end{eqnarray} 
 \begin{eqnarray}   
      \bra{T(k, \varepsilon)} \bar{q}_1  \sigma^{\rho\alpha}\, q_\alpha  b \ket{B(p)}
       & =& - \varepsilon^*_\beta \bigg[\left( g^{\rho\beta} (m_B^2 - m_T^2) - (p + k)^\rho q^\beta\right) T_{2}\nonumber\\&+&  q^\beta \left(q^\rho - \frac{q^2}{m_B^2 - m_T^2} (p + k)^\rho\right) T_{3}\bigg]\,,
     \label{eq:BtoT:tensor6}   
\end{eqnarray} 
where  $\varepsilon_\beta=\frac{\varepsilon_{\beta\alpha}q_{\alpha}}{m_B}$ and $\varepsilon_{\beta\alpha}$ represents the polarization of the tensor meson, which is symmetric with respect to its indices and satisfies the condition $\varepsilon_{\alpha\beta}(k)k^{\beta}=0$. The $\epsilon^{\rho\beta\delta\sigma}$ is the anti-symmetric Levi-Civita tensor, and $\epsilon^{0123}=+1$. Here we emphasize that if $q_1$ represents the strange quark than $B$ stands for $B_s$ meson. In the above-given matrix elements, we represent the momenta of the initial baryon and final tensor meson with $p$ and $k$, respectively, and $q^2=(p-k)^2$ is the momentum transferred square. In these matrices, $V_3$ is related to $V_1$ and $V_2$ as follows:
\begin{eqnarray}
V_3=\frac{m_B+m_T}{2m_T}V_1-\frac{m_B-m_T}{2m_T}V_2.
\label{eq:kinematiccond1}
\end{eqnarray} 
Besides, the unphysical singularities present in the matrix elements given by Eq.~(\ref{eq:BtoT:tensor}) at $q^2=0$  are removed by using the condition
\begin{eqnarray}
V_0(q^2=0)=V_3(q^2=0).
\label{eq:kinematiccond2}
\end{eqnarray}
And also using the algebraic relation between $\sigma_{\mu\nu}$ and $\sigma_{\mu\nu}\gamma_5$, we have
\begin{eqnarray}
T_1(q^2=0)=-T_2(q^2=0).
\label{eq:cond3}
\end{eqnarray}
The summation over the polarization of the tensor meson is performed via
\begin{eqnarray}
\varepsilon_{\mu\nu}(k) \varepsilon^{*}_{\alpha\beta}(k) = \frac{1}{2} \kappa_{\mu\alpha} \kappa_{\nu\beta} + \frac{1}{2} \kappa_{\mu\beta} \kappa_{\nu\alpha} - \frac{1}{3} \kappa_{\mu\nu} \kappa_{\alpha\beta} \, .
\end{eqnarray} 
where
\begin{eqnarray}
\kappa_{\mu\nu} = -g_{\mu\nu} + \frac{k_\mu k_\nu}{m^2_T}.
\end{eqnarray}
Substituting Eqs.~(\ref{Eq:Matrix1})-(\ref{eq:BtoT:tensor6}) into Eq.~(\ref{Eq:corr}) and separating the kinematical structures presented in Table~\ref{tab:Lr.Str.}, we get the hadronic dispersion relation for the seven invariant functions. 
\begin{table}[]
\begin{tabular}{|c|c|c|c|c|c|c|c|c|}
\hline
Lorentz Structure & $\epsilon^{\mu\rho\alpha\beta} q^\nu q_\beta k_\alpha$ & $q^{\nu}g^{\mu\rho}$ & $k^{\rho}q^{\mu}q^{\nu}$ & $q^{\rho}q^{\mu}q^{\nu}$ & $\epsilon^{\mu\rho\alpha\beta} q^\nu q_\beta k_\alpha$  & $k^{\rho}q^{\mu}q^{\nu}$ & $q^{\rho}q^{\mu}q^{\nu}$ \\ \hline \hline
Form Factor       & $A$                                                    & $V_1$                & $V_2$                    & $V_{023}$                & $T_1$                                                  & $T_{23A}$                & $T_{23B}$                \\ \hline
\end{tabular}
\caption{The Lorentz structures used to extract the form factors.}
\label{tab:Lr.Str.}
\end{table}
\begin{eqnarray}
\Pi_i(k^2,q^2)&=&\frac{1}{\pi}\int ds \frac{\textrm{Im}_{k^2}\Pi_i(s,q^2)}{s-k^2}\nonumber\\ 
&=&\frac{A_iF_i(q^2)}{m_T^2-k^2}+\textrm{higher states and continuum contributions},
\end{eqnarray}
where $F_i(q^2)$ is the relevant form factors ($A,~V_1,~V_2,~V_3,~T_1,~T_2$, and $T_3$ ) and $A_i$ are corresponding kinematical factors. The kinematical factors for relevant form factors are given in Appendix~\ref{app:coefficients}. $V_{023}$ combination present in Table~\ref{tab:Lr.Str.} is determined in terms of $V_0,~V_2 $, and $V_3$ as
\begin{eqnarray}
V_{023}=\frac{V_2}{m_B+m_T}+\frac{2m_T(V_3-V_0)}{q^2}.
\end{eqnarray}
And similarly, since $T_2$ and $T_3$ have some common Lorentz structures, we define combined terms, $T_{23A}$ and $T_{23B}$, which are attained from the structures $k^{\rho}q^{\mu}q^{\nu}$ and $q^{\rho}q^{\mu}q^{\nu}$ and from these we obtain the $T_2$ and $T_3$ as
\begin{equation}
\label{eq:T2}
T_2=\frac{2q^2}{m_B^2-m_T^2}T_{23B}+\frac{m_B^2-m_T^2-q^2}{m_B^2-m_T^2}T_{23A},
\end{equation}
and
\begin{equation}
T_3=T_{23A}-2T_{23B}.
\label{eq:T3}
\end{equation}

Now let's calculate the OPE part of the correlation function. The OPE is obtained using the explicit forms of the interpolating currents, Eqs.~(\ref{eq:current1}), (\ref{eq:current2}) and the weak currents with $\Gamma_2^{\rho}$  presented in the Table~\ref{tab:int.current}, into correlation function. In result, we get
\begin{equation}
\label{eq:correlatorOPE2pt} 
    \Pi_{\text{OPE}}^{\mu\nu\rho}(q, k)
        =  \int \text{d}^4 x\,\int  \frac{ \text{d}^4 p'}{(2 \pi)^4} \, e^{i (k - p')\cdot x}\,
        \left[  \Gamma_1^{\mu\nu}\, \frac{\slashed{p}'+m_{q_1}}{m_{q_1}^2 - p'^2}  \Gamma_2^\rho \right]_{\alpha\beta} 
        \bra{0} \bar{q}_{2}^{\alpha}(x) h_{v}^{\beta}(0) \ket{B_{q_2}(v)},
\end{equation}
where $\Gamma_1^{\mu\nu}=\gamma^{\mu}\gamma_5\overleftrightarrow{\mathcal{D}}^{\nu}+\mu\leftrightarrow\nu$, $p'=k-l$ with $l$ being the momentum of the spectator quark present in the $B$ meson. In Eq.~(\ref{eq:correlatorOPE2pt}), the $\alpha,~\beta$ are spinor indices. The nonperturbative matrix elements $ \bra{0} \bar{q}_{2}^{\alpha}(x) h_{v}^{\beta}(0) \ket{B}$ are given in terms of $B$ meson LCDAs and their explicit forms are presented in the Appendix~\ref{app:LCDAs}. 

Before performing the calculation of the OPE part of the correlation function, we would like to note that in the present work, two-particle contributions are taken into account for which the first term in the covariant derivative will be sufficient, and we do not consider its second term, including $A_a^{\nu}$, giving contributions to three particle effects. The relative impact of three-particle contributions to two-particle contributions was analyzed in Refs.~\cite{Gubernari:2018wyi} and it is obtained that it is small. For this reason, we restrict ourselves by considering contributions coming from two-particle DAs.

The OPE side of the correlation function can be written as
\begin{eqnarray}
\label{eq:DecompCorrOPElo-Tensor-VmAC}
%V-ACurrent
\Pi^{\mu\nu\rho}_{\text{(V-A)}}(q, k)  &= &q^\rho q^\mu q^\nu \, \Pi^{(1)}_{\rm OPE}(q^2, k^2)  +  k^\rho q^\mu q^\nu  \, \Pi^{(2)}_{\rm OPE}(q^2, k^2)
 + \epsilon^{\mu\rho\alpha\beta} q^\nu q_\beta k_\alpha \, \Pi^{(3)}_{\rm OPE}(q^2, k^2)  \nonumber\\
 & +&  q^\nu g^{\mu\rho}  \, \Pi^{(4)}_{\rm OPE}(q^2, k^2)  + \cdots \,  , 
\end{eqnarray}
\begin{eqnarray}
\Pi^{\mu\nu\rho}_{\text{Tensor}}(q, k)  &=& q^\rho q^\mu q^\nu \, \Pi^{(5)}_{\rm OPE}(q^2, k^2)  +    k^\rho q^\mu q^\nu  \, \Pi^{(6)}_{\rm OPE}(q^2, k^2)
 +  \epsilon^{\mu\rho\alpha\beta} q^\nu q_\beta k_\alpha \, \Pi^{(7)}_{\rm OPE}(q^2, k^2)  \nonumber\\
 &+& \cdots \, ,
\label{eq:DecompCorrOPElo-Tensor-tensorC}
\end{eqnarray}
where we use the $\cdots$ to represent the other Lorentz structures. By choosing the relevant Lorentz structures and matching their coefficients we get the LCSR for the form factors, (See Table~\ref{tab:Lr.Str.}).

The OPE results for the invariant functions $\Pi_i(k^2,q^2)$ can be written in generic form as
\begin{eqnarray}
\Pi^{(i)}(k^2,q^2)=f_B m_B \sum_k \int dx \frac{I^{(i)}_n (x,q^2)}{(k^2-s(q^2))^n},
\end{eqnarray}
where
\begin{eqnarray}
\label{eq:defs}
    s(x,q^2)=x m^2_B +\frac{m_{q_1}^2-x q^2}{\bar{x}}\,
\end{eqnarray} 
where $\bar{x} = 1 -x$, and  $s'(x,q^2)=\frac{\text{d} s(x,q^2)}{\text{d} x}$. The functions $I^{(i)}_n$ are represented as linear combinations of the four $B$ meson DAs
\begin{eqnarray}
I^{(i)}_n(x,q^2)=\frac{1}{\bar{x}^n}\sum C_{n}^{\psi}(x,q^2)\psi(x m_B),
\end{eqnarray}
where $\psi=\phi_{+},~\bar{\phi},~g_+$, and $g_{-}$  and $x=\frac{\omega}{m_B}$. The coefficient $C_{n}^{\psi}$ are given in Appendix~\ref{app:LCDAs}.

In order to subtract the contributions of higher states and continuum and enhance the contributions of the ground states, we apply the Borel transformation. For this aim, we have used a master formula obtained in~\cite{Aliev:2019ojc} and \cite{Gubernari:2018wyi}, which is as follows:
\begin{eqnarray}
 T&= 
    &\frac{f_B m_B\, }{ N  \, A_{i}} \sum_{n=1}^{\infty}(-1)^{n}\Bigg\{\int_{0}^{x_0} d x \;e^{(-s(x,q^2)+m^2_{T})/M^2} \frac{1}{(n-1)!(M^2)^{n-1}}I_n^{(i)}\nonumber\\
        & +& \Bigg[\frac{1}{(n-1)!}e^{(-s(x,q^2)+m^2_{T})/M^2}\sum_{j=1}^{n-1}\frac{1}{(M^2)^{n-j-1}}\frac{1}{s'}
        \left(\frac{\text{d}}{\text{d}x}\frac{1}{s'}\right)^{j-1}I_n^{(i)}\Bigg]_{x=x_0}\Bigg\rbrace\,,
        \label{eq:masterformula}
\end{eqnarray}
where $N=\sqrt{2}$ for $f_2$ and $a_2$ meson states and $N=1$ for other states and the differential operator acts as $\left(\frac{\text{d}}{\text{d}x}\frac{1}{s'}\right)^{n} I(x) \to 
    \left(\frac{\text{d}}{\text{d}x}\frac{1}{s'}\left(\frac{\text{d}}{\text{d}x}\frac{1}{s'}\dots I(x)\right)\right)$. 
The quantity $x_0$ is the solution of the equation $s(x,q^2)=s_0$, which leads to
\begin{eqnarray}
x_0 = \frac{s_0+m_B^2-q^2- \sqrt{4 (m_{q_1}^2-s_0) m_B^2 + (m_B^2 +s_0 -q^2)^2 }}{2 m_B^2}\, .    
\end{eqnarray}
From our analyses we obtain the form factors, $T=A,~V_1~,V_2,~V_{30},~T_1,~T_{23A}$ and $T_{23B}$ directly, and the remaining ones, $A_0,~T_2$ and $T_3$ are obtained by using
\begin{eqnarray}
A_0=A_3-A_{30},
\end{eqnarray} 
and Eqs.~(\ref{eq:T2}) and (\ref{eq:T3}).

Note that in the derivation of Eq.~(\ref{eq:masterformula}) we have used narrow-width approximation; i.e. the widths of the tensor mesons are taken zero.

At the end of this section we present the expression for the decay widths of the considered transitions. In the Standard Model the $B_{(s)}\rightarrow T l^+l^-$ transitions are governed by the effective Hamiltonian given as:
\begin{eqnarray}
\mathcal{H}_{eff}=-\frac{4G_{F}}{\sqrt{2}}V_{tb}^{\ast }V_{tq_1}{\sum\limits_{i=1}^{10}}
C_{i}({\mu }) \mathcal{O}_{i}({\mu }),
\label{eq:effectiveH}
\end{eqnarray}
where $\mathcal{O}_{i}({\mu })$ and $C_{i}({\mu })$  are the effective operators and 
the corresponding Wilson coefficients at the renormalization scale $\mu$, respectively. There are ten operators, $\mathcal{O}_{i}({\mu })$ in \refeq{effectiveH}, and only the $\mathcal{O}_{7}$, $\mathcal{O}_{9}$ and $\mathcal{O}_{10}$ contribute  to the $B_{(s)}\rightarrow T l^+l^-$ transitions. These operators are given explicitly as
\begin{eqnarray}
\mathcal{O}_{7} &=& \frac{e^{2}}{16\pi ^{2}}m_{b}( \bar{q_1}\sigma _{\mu \nu
}\frac{(1+\gamma_5)}{2}b) F^{\mu \nu },\,   \nonumber \\
\mathcal{O}_{9} &=& \frac{e^{2}}{16\pi ^{2}}(\bar{q_1}\gamma _{\mu }\frac{(1-\gamma_5)}{2}b)(\bar{\ell}\gamma
^{\mu } \ell),\, \nonumber\\  
\mathcal{O}_{10} &=& \frac{e^{2}}{16\pi ^{2}}(\bar{q_1}\gamma _{\mu }\frac{(1-\gamma_5)}{2} b)(\bar{\ell}
\gamma ^{\mu }\gamma _{5} \ell).  
\end{eqnarray}
The Wilson coefficients are given in Refs.~\cite{Buras:1994dj,Buras:2011we,Hurth:2010tk}. After standard calculations for the decay widths for considered transitions, we get
\begin{eqnarray}
\frac{d\Gamma }{dq^{2}} &=& \frac{G_{F}^{2}\alpha ^{2}}{2^{11}\pi
^{5}m_{B}^{3}}\left\vert V_{tb}V_{tq_1}^{\ast }\right\vert ^{2} \sqrt{\lambda\left(1-\frac{%
4m_{\ell}^{2}}{q^{2}}\right)} \left(
\frac{\left\vert \tilde{F}_1 \right\vert ^{2}\left( 2m_{\ell}^{2}+q^{2}\right)
\lambda ^{2}}{6m_{B}^{2}m_{T}^{2}}\right.  \nonumber \\
&+& \frac{\left\vert \tilde{F}_2 %
\right\vert ^{2}m_{B}^{2}\left( 2m_{\ell}^{2}+q^{2}\right) \left(
10q^{2}m_{T}^{2}+\lambda \right) \lambda }{9m_{T}^{4}q^{2}} +\frac{\left\vert \tilde{F}_3 \right\vert ^{2}\left(
2m_{\ell}^{2}+q^{2}\right) \lambda ^{3}}{9m_{B}^{2}m_{T}^{4}q^{2}}-%
\frac{\left\vert \tilde{F}_4 \right\vert ^{2}\left( 4m_{\ell}^{2}-q^{2}\right)
\lambda ^{2}}{6m_{B}^{2}m_{T}^{2}}\nonumber \\
&+& \frac{\left\vert \tilde{F}_5 %
\right\vert ^{2}m_{B}^{2}\left( 2\left( \lambda -20m_{T}^{2}q^{2}\right) m_{\ell}^{2}+q^{2}\left( 10q^{2}m_{T}^{2}+\lambda
\right) \right) \lambda }{9m_{T}^{4}q^{2}} +\frac{%
2\left\vert \tilde{F}_7 \right\vert ^{2}m_{\ell}^{2}q^{2}\lambda ^{2}}{%
3m_{B}^{2}m_{T}^{4}}  \nonumber\\
&+& \frac{\left\vert \tilde{F}_6 \right\vert ^{2}\left( 2\left( \left(
m_{B}^{2}-m_{T}^{2}\right) ^{2}-2q^{4}+4\left(
m_{B}^{2}+m_{T}^{2}\right) q^{2}\right) m_{\ell}^{2}+q^{2}\lambda
\right) \lambda ^{2}}{9m_{B}^{2}m_{T}^{4}q^{2}}\nonumber \\
&+& \frac{2\mbox{Re} (\tilde{F}_2 \tilde{F}_3 ^{\ast}) \left( 2m_{\ell}^{2}+q^{2}\right) \left(
-m_{B}^{2}+m_{T}^{2}+q^{2}\right) \lambda ^{2}}{9m_{T}^{4}q^{2}}+\frac{4\mbox{Re} ( \tilde{F}_6 \tilde{F}_7 ^{\ast} )m_{\ell}^{2}\left(
m_{B}^{2}-m_{T}^{2}\right) \lambda ^{2}}{3m_{B}^{2}m_{T}^{4}}  \notag \\
&+& \frac{2\mbox{Re} (  \tilde{F}_5 \tilde{F}_6 ^{\ast} )\left( q^{2}\left(
-m_{B}^{2}+m_{T}^{2}+q^{2}\right) -2m_{\ell}^{2}\left(
m_{B}^{2}-m_{T}^{2}+2q^{2}\right) \right) \lambda ^{2}}{%
9m_{T}^{4}q^{2}}\nonumber \\
& -&\left.\frac{4\mbox{Re} (  \tilde{F}_5 \tilde{F}_7 ^{\ast}  )m_{\ell}^{2}\lambda ^{2}}{3m_{T}^{4}}\right).  
\label{eq:BtoTllDecay}
\end{eqnarray} 
In Eq.~(\ref{eq:BtoTllDecay}) the $\lambda$ means $\lambda(m_B^2,m_T^2,q^2)$ and is defined as
\begin{eqnarray}
\lambda(x,y,z)=x^2+y^2+z^2-2xy-2xz-2yz,
\end{eqnarray}
and $\tilde{F}_i$ are
\begin{eqnarray}
\tilde{F}_1 &=& - C_{9}^{eff}(\mu )\frac{2}{m_{B}+m_{T}}%
{A}(q^{2})-C_{7}^{eff}(\mu )\frac{4m_{b}}{q^{2}}{T}_{1}\left( q^{2}\right)
\nonumber \\
\tilde{F}_2 &=& -\frac{\left( m_{B}+m_{T}\right) }{m_{B}^{2}}\left[
C_{9}^{eff}(\mu ){V}_{1}(q^{2})+C_{7}^{eff}(\mu )\frac{2m_{b}(m_{B}-m_{T})}{q^{2}}{ T}_{2}\left( q^{2}\right) \right]  \nonumber \\
\tilde{F}_3 &=& -C_{9}^{eff}(\mu )\frac{1}{m_{B}+m_{T}}%
{V}_{2}(q^{2})-C_{7}^{eff}(\mu )\frac{2m_{b}}{q^{2}}\left[ {T}_{2}\left(
q^{2}\right) +\frac{q^{2}}{m_{B}^{2}-m_{T}^{2}}{T}_{3}\left(
q^{2}\right) \right]  \nonumber \\
\tilde{F}_4 &=& - C_{10}\frac{2}{m_{B}+m_{T}}{A}(q^{2})  \nonumber
\\
\tilde{F}_5 &=& -C_{10}\frac{(m_{B}+m_{T})}{m_{B}^{2}}%
{ V}_{1}(q^{2})  \nonumber \\
\tilde{F}_6 &=& -C_{10}\frac{1}{m_{B}+m_{T}}{V}_{2}(q^{2})
\notag \\
\tilde{F}_7 &=& C_{10}\left[ \frac{2m_{T}}{q^{2}}%
{V}_{0}(q^{2}) - \frac{(m_{B}+m_{T})}{q^{2}}{V}_{1}(q^{2})+\frac{%
(m_{B}-m_{T})}{q^{2}}{V}_{2}(q^{2})\right] .
\label{eq:DW_func}
\end{eqnarray}

\section{\label{III} Numerical analyses}

In this section, we present the numerical analyses of the LCSR for the form factors derived in Section~\ref{II}. To get the numerical results, we need input parameters, such as decay constants, the masses of the considered mesons, and the masses of the quarks present in the considered states. The masses and decay constants of the tensor mesons are presented in Table~\ref{tab:inputs}.
\begin{table}[]
\begin{tabular}{|c|c|c|c|c|}
\hline
State    & $m~(\textrm{GeV})$~\cite{Aliev:2017apq} & Decay constants~\cite{Aliev:2017apq} & $M^2~(\textrm{GeV}^2)$ & $s_0~(\textrm{GeV}^2)$ \\ \hline \hline
$ K_2$   & $1.85\pm0.14$                                   & $(6.2\pm 0.4)\times10^{-2}$                          & $1.5-2.5$            & $2.1^2-2.5^2$        \\ \hline
$\phi_2$ & $2.00\pm0.16$                                   & $(7.7\pm 0.1)\times10^{-2}$                          & $1.5-2.5$            & $2.1^2-2.5^2$        \\ \hline
$f_2$    & $1.78\pm0.12$                                   & $(7.4\pm 0.1)\times10^{-2}$                          & $1.5-2.5$            & $2.0^2-2.3^2$        \\ \hline
$ a_2$   & $1.78\pm0.12$                                   & $(7.4\pm 0.1)\times10^{-2}$                          & $1.5-2.5$            & $2.0^2-2.3^2$        \\ \hline
\end{tabular}
\caption{The input masses and decay constants for the tensor mesons~\cite{Aliev:2017apq} used in the analyses of the decays, and Borel parameters, $M^2$ and effective threshold parameters, $s_0$ used for each decay in the present work.}
\label{tab:inputs}
\end{table} 
The masses  and the decay constants of the $B$ mesons are taken as $m_B=5279.72\pm 0.08$~MeV~\cite{ParticleDataGroup:2024cfk}, $m_{\bar{B}_s^0}=5366.93\pm 0.10$~MeV~\cite{ParticleDataGroup:2024cfk} and $f_B=189.4\pm1.4$~MeV~\cite{Bazavov:2017lyh} and $f_{B_s}=230.7\pm1.3$~MeV~\cite{Bazavov:2017lyh}. The masses of the $b$ and $s$ quarsks are $m_b(m_b)=4.18$~GeV and $m_s(1~\textrm{GeV})=0.128$~GeV~\cite{ParticleDataGroup:2024cfk}. The nonperturbative parameters in the LCDAs of $B$ meson are used as $\lambda_B=360\pm110$~MeV~\cite{Janowski:2021yvz}. The values of 
$\lambda_E^2$ and $\lambda_H^2$ are obtained by averaging central values in Refs.~\cite{Rahimi:2020zzo,Nishikawa:2011qk}  $\lambda_E^2=0.02\pm0.03~\textrm{GeV}^2$ and $\lambda_H^2=0.11\pm0.08~\textrm{GeV}^2$~\cite{Rahimi:2020zzo,Nishikawa:2011qk} (see also \cite{Braun:2003wx}). For the value of the $\lambda_{B_s}$ we have used the result of Ref.~\cite{Khodjamirian:2020hob} $\lambda_{B_s}/\lambda_B=1.19\pm 0.14$.

Besides, we also have two auxiliary parameters entering the sum rules for the form factors, namely, the Borel parameter, $M^2$, and continuum threshold, $s_0$. Their proper values are determined following the criteria: The working range of the Borel parameter $M^2$ is determined by ensuring that the contributions from higher-twist power corrections and continuum states are significantly suppressed compared to the leading twist terms. The lower bound of $M^2 $ is set to ensure that higher-twist contributions should be suppressed. On the other hand, the upper bound is limited by the requirement that the continuum contributions, which are modeled through quark-hadron duality, do not dominate the sum rule. This careful adjustment of $M^2$ enables an accurate and stable extraction of physical observables while maintaining the validity of the operator product expansion. To fix the threshold parameter, following the works of Ref.~\cite{SentitemsuImsong:2014plu,Ball:2004ye,Jamin:2001fw,Bharucha:2015bzk} we have used the mass sum rules and demand that the mass sum rules reproduce the experimental mass values of the considered tensor mesons within $\pm5\%$. The resultant working intervals for Borel parameters and threshold parameters are also presented in Table~\ref{tab:inputs}.  

\begin{figure} []
\centering
\begin{tabular}{cccc}
\includegraphics[width=0.4\textwidth]{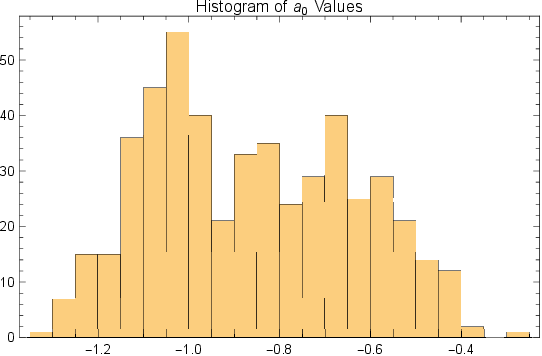} &
\includegraphics[width=0.4\textwidth]{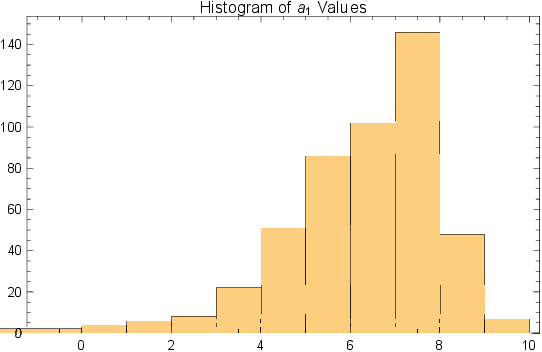} &
%\includegraphics[width=0.3\textwidth]{grV1phi.eps} \\
%\textbf{(a)}  & \textbf{(b)} & \textbf{(c)}  \\[6pt]
\end{tabular}
\caption{ The histograms of the fit parameters $a_0$ and $a_1$ of the form factor $A$ obtained for $B \rightarrow K_2$ transition. 
%\textbf{(a)} Some text
%\textbf{(b)} Some text
%\textbf{(c)} Some text
%\textbf{(d)} Some text
%\textbf{(e)} Some text
}
\label{fig:histogram}
\end{figure}

With the determined Borel mass parameter $M^2$, and threshold parameter $s_0$, we can determine the LCSR results for the form factors. It should be noted that the results of the LCSRs are only reliable up to a value of the $q^2 < 0$. It is justified by the fact that at $q^2\geq 0$ the subleading twist contributions give the same order of magnitude as the leading twist ones at $q^2=0$. Therefore, to get the results of the form factors in the physical region, we need to parameterize the results of LCSRs using a simple pole fit function. To this end, we use a $z$-series expansion as suggested in~\cite{Bharucha:2015bzk}: 
\begin{eqnarray}
F(q^2) = \frac{1}{1 - \frac{q^2} { m_{fit}^2}}\, \sum_{n=0}^{1} a_n \left[z(q^2) - z(0)\right]^n\,,
\label{eq:fitform}
\end{eqnarray}        
where
\begin{equation}
z(s) = \frac{\sqrt{t_+ - s} - \sqrt{t_+ - t_0}}{\sqrt{t_+ - s} + \sqrt{t_+ - t_0}}, 
\end{equation}
with $t_\pm = (m_B \pm m_{T})^2$ and $t_0 = t_+ \left(1 - \sqrt{1 - \frac{t_-} { t_+}}\right)$.
In the Eq.~(\ref{eq:fitform}), $a_n=a_0,~a_1$ are the fit parameters, and they are  given in the Table~\ref{tab:fitparameters} for each form factor, and $m_{fit}$ are the pole masses with values given also in the Table~\ref{tab:fitparameters}. We demand that this parameterization of the form factors should reproduce accurately the LCSR results in $q^2<0$ domain. 
\begin{table}[]
\begin{tabular}{|c|c|c|c|c|}
\hline
Final tensor state                   & Form factor & $a_0$          & $a_1$   & $m_{fit}~(\textrm{GeV})$~\cite{Gubernari:2018wyi}\\ \hline \hline
\multirow{7}{*}{$ K_2$}   & $A$         & $-0.873\pm 0.220$   & $5.770 \pm 3.050$  & $5.829$                                             \\ \cline{2-5} 
                          & $V_0$       & $-0.420 \pm 0.050$  & $-6.788\pm 0.343$   & $5.711$                                             \\ \cline{2-5} 
                          & $V_1$       & $-0.459\pm 0.107$   & $-0.216\pm 0.710$  & $5.412$                                             \\ \cline{2-5} 
                          & $V_2$       & $-0.545\pm 0.223$   & $ 2.563\pm 3.300$  & $5.412$                                             \\ \cline{2-5} 
                          & $T_1$       & $-0.567\pm 0.149$   & $3.645\pm 1.845$   & $5.829$                                             \\ \cline{2-5} 
                          & $T_2$       & $0.599\pm 0.151$    & $-1.215\pm 3.156$  & $5.412$                                             \\ \cline{2-5} 
                          & $T_3$       & $0.303\pm 0.283$    & $-1.780\pm 4.248$  & $5.412$                                             \\ \hline \hline
\multirow{7}{*}{$\phi_2$} & $A$         & $-0.883\pm 0.301$   & $5.903\pm 1.665$   & $5.829$                                             \\ \cline{2-5} 
                          & $V_0$       & $-0.424\pm 0.068 $  & $-6.571\pm 0.612$  & $5.711$                                             \\ \cline{2-5} 
                          & $V_1$       & $-0.513\pm 0.142$   & $0.695\pm 0.816$   & $5.412$                                             \\ \cline{2-5} 
                          & $V_2$       & $-0.615\pm 0.270$   & $3.594\pm 1.683$   & $5.412$                                             \\ \cline{2-5} 
                          & $T_1$       & $-0.598\pm 0.184$   & $3.998\pm 1.063$  & $5.829$                                             \\ \cline{2-5} 
                          & $T_2$       & $0.648\pm 0.223$    & $-1.986\pm 1.512$ & $5.412$                                             \\ \cline{2-5} 
                          & $T_3$       & $0.304\pm 0.347$    & $-2.208\pm 2.587$ & $5.412$                                             \\ \hline \hline
\multirow{7}{*}{$f_2$}    & $A$         & $-0.499\pm 0.147$   & $3.688\pm 0.919$  & $5.724$                                             \\ \cline{2-5} 
                          & $V_0$       & $-0.261\pm 0.027$   & $-2.895\pm 0.244$ & $5.681$                                             \\ \cline{2-5} 
                          & $V_1$       & $-0.272\pm 0.077$   & $0.455\pm 0.515$  & $5.325$                                             \\ \cline{2-5} 
                          & $V_2$       & $-0.317\pm 0.151$   & $ 1.976\pm 1.017$ & $5.325$                                             \\ \cline{2-5} 
                          & $T_1$       & $-0.341\pm 0.099$   & $2.437\pm 0.588$  & $5.724$                                             \\ \cline{2-5} 
                          & $T_2$       & $0.356\pm 0.125$    & $-1.351\pm 0.546$ & $5.325$                                             \\ \cline{2-5} 
                          & $T_3$       & $0.143\pm 0.194$   & $-1.167\pm1.623$   & $5.325$                                             \\ \hline \hline
\multirow{7}{*}{$ a_2$}   & $A$         & $-0.503\pm 0.151$   & $3.666\pm 0.952$  & $5.724$                                             \\ \cline{2-5} 
                          & $V_0$       & $-0.248\pm 0.024$   & $-2.762\pm 0.202$ & $5.681$                                             \\ \cline{2-5} 
                          & $V_1$       & $-0.269\pm 0.076$   & $0.450\pm 0.553$  & $5.325$                                             \\ \cline{2-5} 
                          & $V_2$       & $-0.319\pm 0.149$   & $2.028\pm 1.025$  & $5.325$                                             \\ \cline{2-5} 
                          & $T_1$       & $-0.342\pm 0.104$   & $2.401\pm 0.610$  & $5.724$                                             \\ \cline{2-5} 
                          & $T_2$       & $ 0.342\pm 0.108$   & $-1.379\pm 0.796$ & $5.325$                                             \\ \cline{2-5} 
                          & $T_3$       & $0.155\pm 0.187$    & $ -1.349\pm 1.643$ & $5.325$                                             \\ \hline
\end{tabular}
\caption{The results obtained for fit parameters $a_0$ and $a_1$ and the values of the $m_{fit}$~\cite{Gubernari:2018wyi} used in the fit functions of each decay.}
\label{tab:fitparameters}
\end{table}
Fit parameters obtained in our analyses satisfy the conditions imposed by the kinematical conditions given in the Eqs.~(\ref{eq:kinematiccond2}) and (\ref{eq:cond3}).  That means the $a_0$ values in the fit functions of $T_1$ and $T_2$ for each transition are equal, and relevant $a_0$ in the fit functions of each transition must satisfy the Eq.~(\ref{eq:kinematiccond2}), when corresponding $a_0$ values of the $V_0$, $V_1$, and $V_2$ fit functions are used in the places of $V_0$ in Eq.~(\ref{eq:kinematiccond2}), and $V_1$, and $V_2$ in the Eq.~(\ref{eq:kinematiccond1}). The errors in the fit parameters given in Table~\ref{tab:fitparameters} arise from the uncertainties present in the input parameters and the auxiliary parameters $s_0$ and $M^2$ used in the analyses.To get these uncertainties corresponding to the fit parameters of each form factor, we conduct a Monte Carlo analyses by randomly sampling 500 values for each variable within their specified ranges. We calculate the form factors for each set of random samples and visualized the resulting distributions of the fit parameters $a_0$ and $a_1$ using histograms. From these histograms, we extracted the mean values and standard deviations of these parameters. Via this application we estimate both the central values of the fit parameters and their associated uncertainties. To provide an example of these uncertainty analyses, we provide the histograms Fig.~\ref{fig:histogram} corresponding to the fit parameters of the form factor $A$ of $B \rightarrow K_2$ transition. The similar analyses was performed for all remaining form factors. The results are collected in Table~\ref{tab:fitparameters}.

\begin{figure} []
\centering
\begin{tabular}{cccc}
\includegraphics[width=0.3\textwidth]{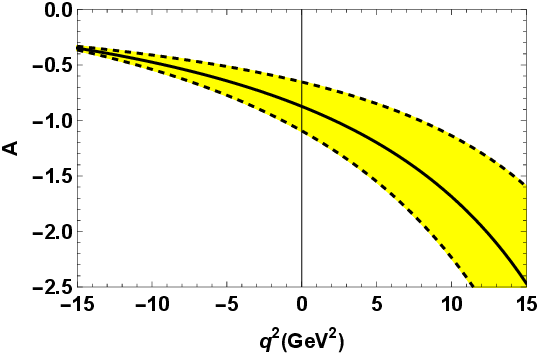} &
\includegraphics[width=0.3\textwidth]{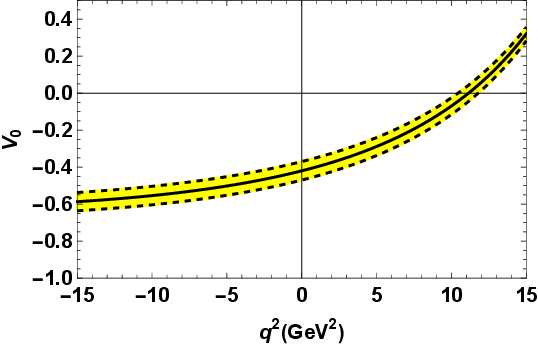} &
\includegraphics[width=0.3\textwidth]{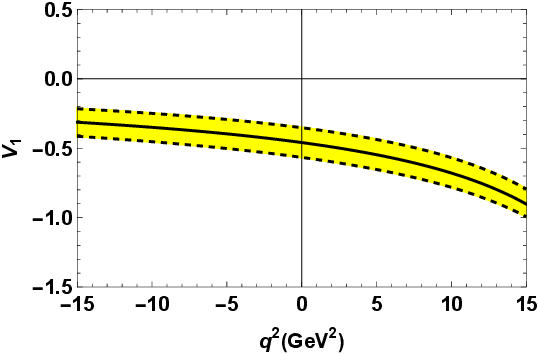} \\
%\textbf{(a)}  & \textbf{(b)} & \textbf{(c)}  \\[6pt]
\end{tabular}
\begin{tabular}{cccc}
\includegraphics[width=0.3\textwidth]{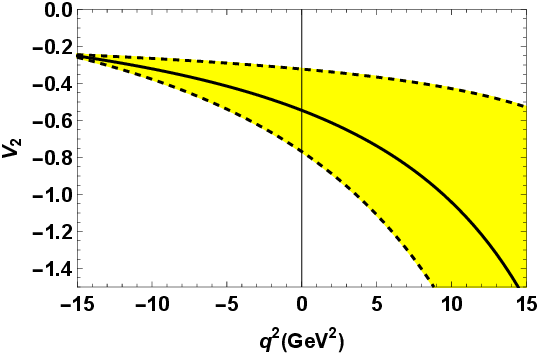} &
\includegraphics[width=0.3\textwidth]{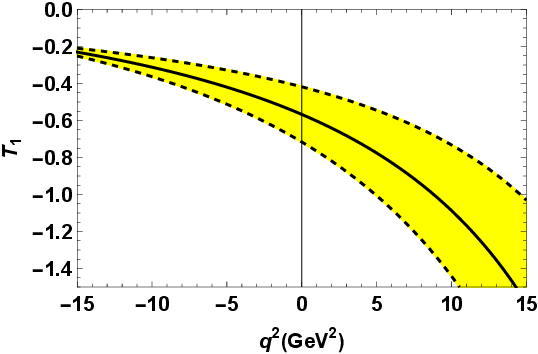} &
\includegraphics[width=0.3\textwidth]{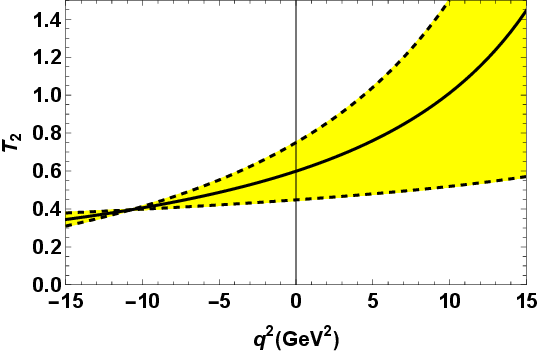} \\
%\textbf{(d)}  & \textbf{(e)}  \\[6pt]
\end{tabular}
\begin{tabular}{cccc}
\includegraphics[width=0.3\textwidth]{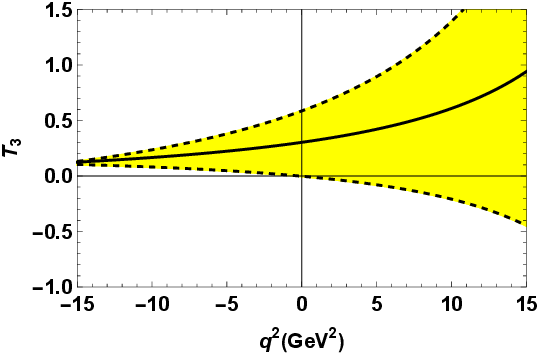} \\
%\textbf{(d)}  & \textbf{(e)}  \\[6pt]
\end{tabular}
\caption{ The variation of the form factors of $B \rightarrow K_2$ transition as a function of $q^2$ at different values of $M^2$ and $s_0$ obtained from the fits attained from the LCSRs results. The error margins arising from the variation of the input parameters are represented by the yellow bands.
%\textbf{(a)} Some text
%\textbf{(b)} Some text
%\textbf{(c)} Some text
%\textbf{(d)} Some text
%\textbf{(e)} Some text
}
\label{fig:K2}
\end{figure}

\begin{figure} []
\centering
\begin{tabular}{cccc}
\includegraphics[width=0.3\textwidth]{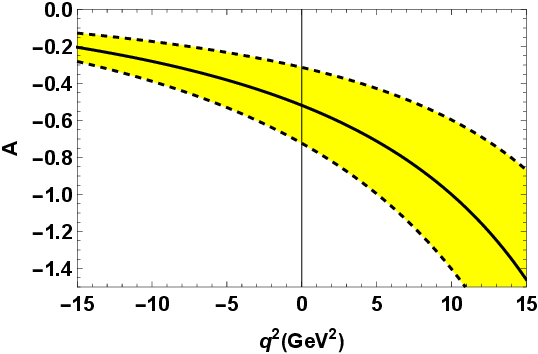} &
\includegraphics[width=0.3\textwidth]{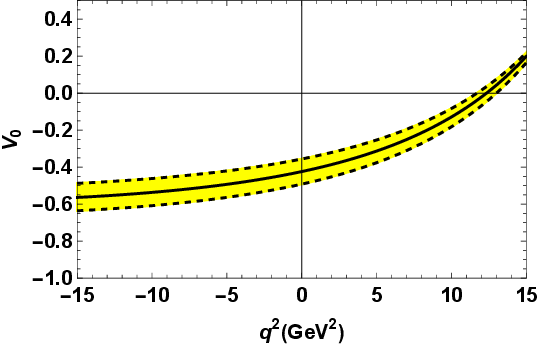} &
\includegraphics[width=0.3\textwidth]{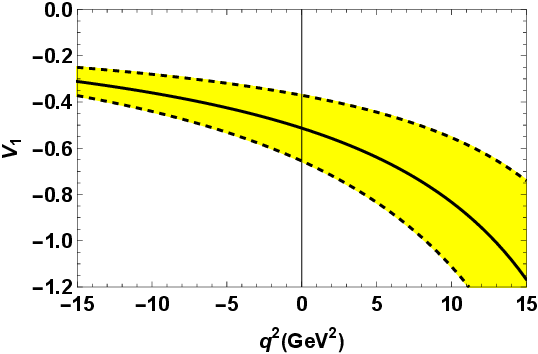} \\
%\textbf{(a)}  & \textbf{(b)} & \textbf{(c)}  \\[6pt]
\end{tabular}
\begin{tabular}{cccc}
\includegraphics[width=0.3\textwidth]{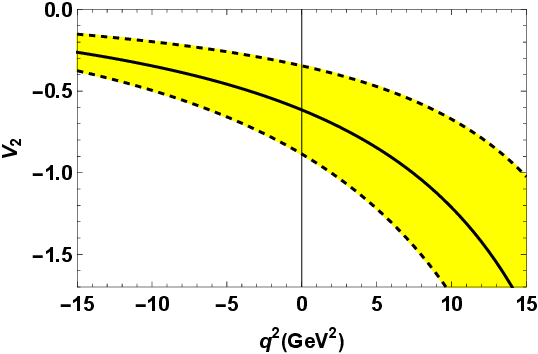} &
\includegraphics[width=0.3\textwidth]{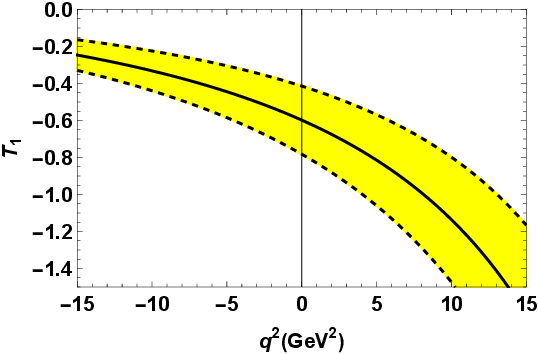} &
\includegraphics[width=0.3\textwidth]{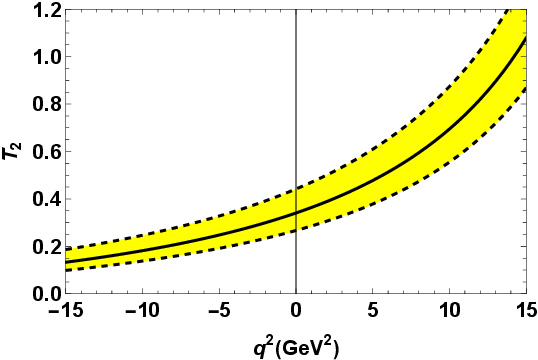} \\
%\textbf{(d)}  & \textbf{(e)}  \\[6pt]
\end{tabular}
\begin{tabular}{cccc}
\includegraphics[width=0.3\textwidth]{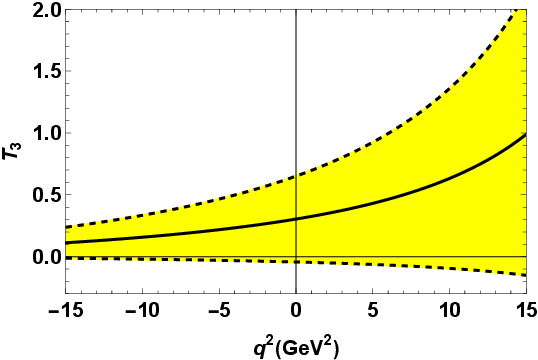} \\
%\textbf{(d)}  & \textbf{(e)}  \\[6pt]
\end{tabular}
\caption{ The variation of the form factors of the $B_s \rightarrow \phi_2$ transition as a function of $q^2$ at different values of $M^2$ and $s_0$ obtained from the fits attained from the LCSRs results. The error margins arising from the variation of the input parameters are represented by the yellow bands. 
%\textbf{(a)} Some text
%\textbf{(b)} Some text
%\textbf{(c)} Some text
%\textbf{(d)} Some text
%\textbf{(e)} Some text
}
\label{fig:phi}
\end{figure}

\begin{figure} []
\centering
\begin{tabular}{cccc}
\includegraphics[width=0.3\textwidth]{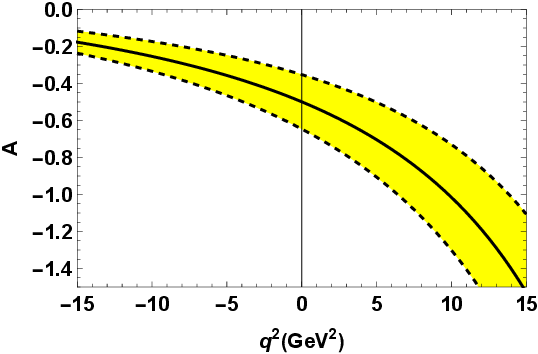} &
\includegraphics[width=0.3\textwidth]{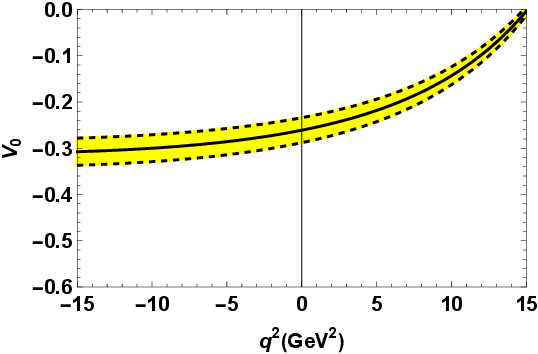} &
\includegraphics[width=0.3\textwidth]{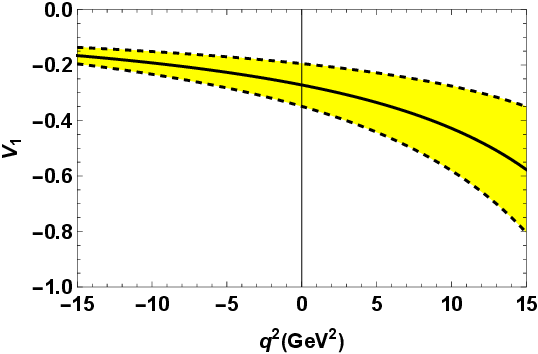} \\
%\textbf{(a)}  & \textbf{(b)} & \textbf{(c)}  \\[6pt]
\end{tabular}
\begin{tabular}{cccc}
\includegraphics[width=0.3\textwidth]{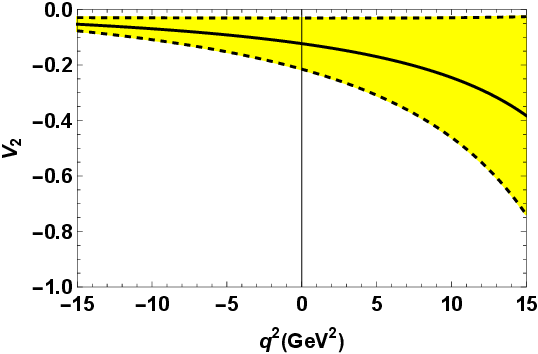} &
\includegraphics[width=0.3\textwidth]{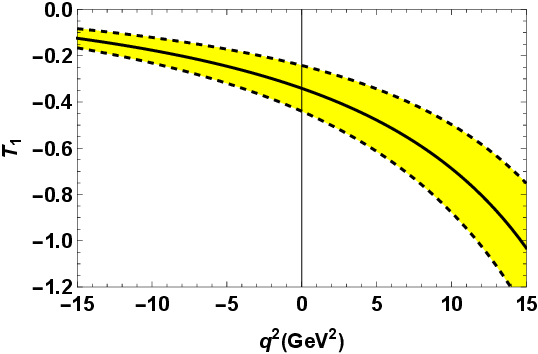} &
\includegraphics[width=0.3\textwidth]{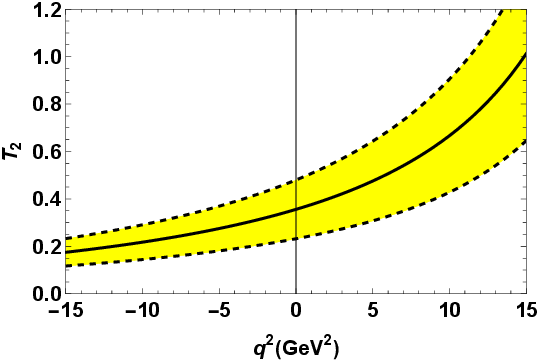} \\
%\textbf{(d)}  & \textbf{(e)}  \\[6pt]
\end{tabular}
\begin{tabular}{cccc}
\includegraphics[width=0.3\textwidth]{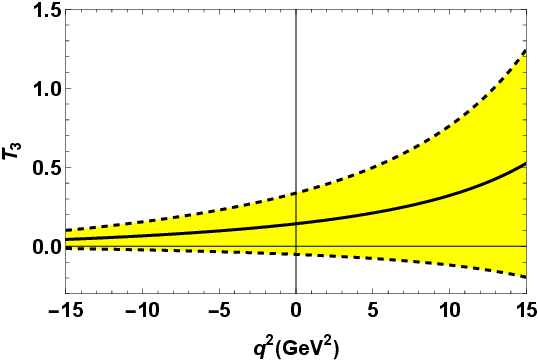} \\
%\textbf{(d)}  & \textbf{(e)}  \\[6pt]
\end{tabular}
\caption{ The variation of the form factors of the $B \rightarrow f_2$ transition as a function of $q^2$ at different values of $M^2$ and $s_0$ obtained from the fits attained from the LCSRs results. The error margins arising from the variation of the input parameters are represented by the yellow bands. 
%\textbf{(a)} Some text
%\textbf{(b)} Some text
%\textbf{(c)} Some text
%\textbf{(d)} Some text
%\textbf{(e)} Some text
}
\label{fig:f2}
\end{figure}

\begin{figure} []
\centering
\begin{tabular}{cccc}
\includegraphics[width=0.3\textwidth]{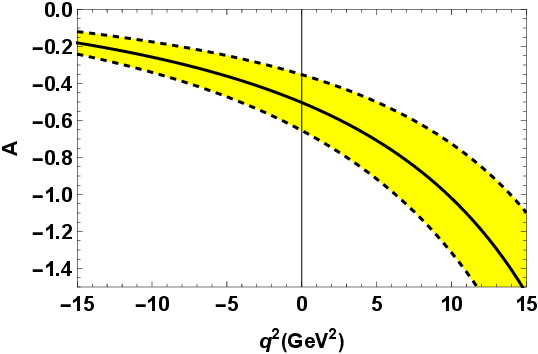} &
\includegraphics[width=0.3\textwidth]{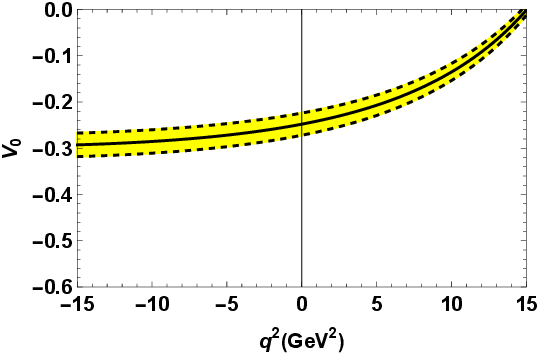} &
\includegraphics[width=0.3\textwidth]{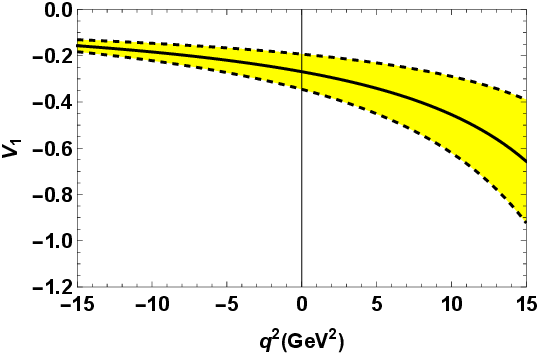} \\
%\textbf{(a)}  & \textbf{(b)} & \textbf{(c)}  \\[6pt]
\end{tabular}
\begin{tabular}{cccc}
\includegraphics[width=0.3\textwidth]{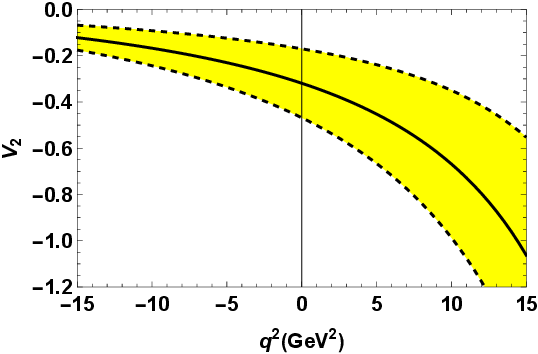} &
\includegraphics[width=0.3\textwidth]{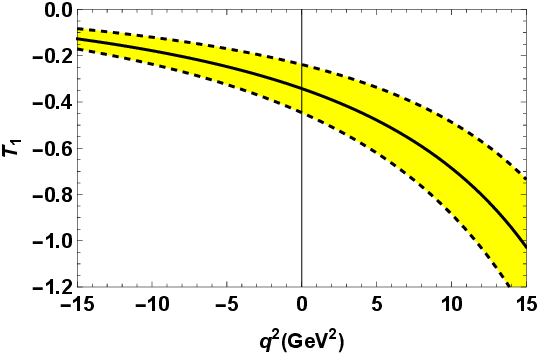} &
\includegraphics[width=0.3\textwidth]{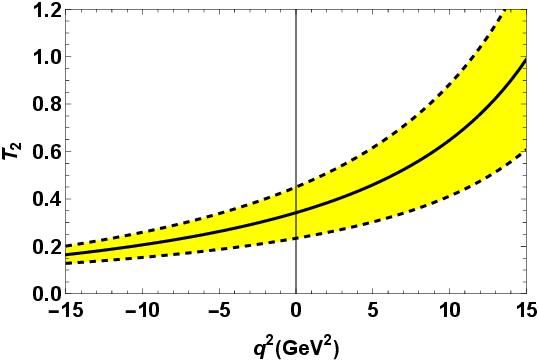} \\
%\textbf{(d)}  & \textbf{(e)}  \\[6pt]
\end{tabular}
\begin{tabular}{cccc}
\includegraphics[width=0.3\textwidth]{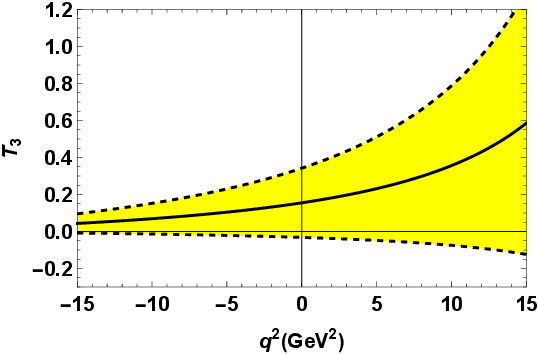} \\
%\textbf{(d)}  & \textbf{(e)}  \\[6pt]
\end{tabular}
\caption{ The variation of the form factors of the $B \rightarrow a_2$ transition as a function of $q^2$ at different values of $M^2$ and $s_0$ obtained from the fits attained from the LCSRs results. The error margins arising from the variation of the input parameters are represented by the yellow bands. 
%\textbf{(a)} Some text
%\textbf{(b)} Some text
%\textbf{(c)} Some text
%\textbf{(d)} Some text
%\textbf{(e)} Some text
}
\label{fig:a2}
\end{figure}

To illustrate their $q^2$ dependence, the graphs of form factors for each transition with the variation of $q^2$ are presented in Figs.~\ref{fig:K2}-\ref{fig:a2} using the corresponding fits obtained from the LCSR analyses. The yellow bands show the uncertainties entering the calculations with the variations of the input parameters.  

After getting the form factors, we use these results in the calculations of the widths of the corresponding decays using the Eq.~(\ref{eq:BtoTllDecay}). The numerical values of the parameters present in the Eq.~(\ref{eq:BtoTllDecay}) are: $G_F=1.167\times 10^{-5}~\mathrm{GeV}^2$, $C_7^{eff}(m_b)=-0.306$~\cite{Asatrian:2016meg}, $C_9^{eff}(m_b)=4.344$~\cite{Buras:1994dj,Asatrian:2016meg}, and $C_{10}=-4.669$~\cite{Buras:1994dj,Asatrian:2016meg}, $V_{tb}=0.77^{+0.18}_{-0.24}$, $V_{ts}=0.0406\pm 0.0027$, and $V_{td}=(8.1\pm 0.5)\times10^{-3}$~~\cite{ParticleDataGroup:2024cfk}. Using the input parameters and the results of LCSR calculations for the form factors, we obtain the branching ratios for the considered decays $B_{(s)}\rightarrow Tl^+l^-$ and the results are present in the Table~\ref{tab:BR-table}.
\begin{table}[]
\begin{tabular}{|c|c|}
\hline
Decay                                  & $BR$                           \\ \hline \hline
$B \rightarrow K_{2}e^+e^-$            & $(9.68\pm 4.22)\times 10^{-7}$ \\ \hline
$B\rightarrow K_{2}\mu^+\mu^-$         & $(5.63\pm 3.06)\times 10^{-7}$ \\ \hline
$B_{s}\rightarrow \phi_2 e^+e^-$       & $(1.43\pm 0.70)\times 10^{-6}$ \\ \hline
$B_{s}\rightarrow \phi_2 \mu^+\mu^-$   & $(8.98\pm 4.80)\times 10^{-7}$ \\ \hline
$B \rightarrow f_2 e^+e^-$             & $(2.40\pm 1.20)\times 10^{-8}$ \\ \hline
$B \rightarrow f_2 \mu^+\mu^-$         & $(1.54\pm 0.70)\times 10^{-8}$ \\ \hline
$B \rightarrow a_2 e^+e^-$             & $(2.31\pm 1.05)\times 10^{-8}$ \\ \hline
$B \rightarrow a_2 \mu^+\mu^-$         & $(1.50\pm 0.72)\times 10^{-8}$ \\ \hline
\end{tabular}
\caption{The branching ratios, $BRs$, obtained for each decay.}
\label{tab:BR-table}
\end{table}

From Table\ref{tab:BR-table} we see that branching ratios for $B \rightarrow T l^+l^-$ decays are  at the $10^{-7}\div 10^{-8}$ order and can quite detectable, especially at LHCb and KEK. Besides in Ref.~\cite{Aliev:2019ojc} the BRs for the $B \rightarrow T l^+ l^-$ transition for $K_2$ tensor meson with $J^P=2^+$ was obtained as $BR(B \rightarrow K_2 e^+ e^-)=(7.72\pm 4.28)\times 10^{-7}$ and. $BR(B \rightarrow K_2 \mu^+ \mu^-)=(6.05\pm 3.81)\times 10^{-7}$. The $B \rightarrow T l^+ l^-$ transitions for $K_2,~a_2$, and $f_2$ tensor meson with $J^P=2^+$ were also considered in Ref.~\cite{Zuo:2021kui} whose results were: $BR(B \rightarrow K_2 e^+ e^-)=(3.90^{+ 0.64}_{-0.44}{}^{+0.69}_{-0.61})\times 10^{-7}$, $BR(B \rightarrow K_2 \mu^+ \mu^-)=(2.27^{+ 0.37}_{-0.26}{}^{+0.40}_{-0.35})\times 10^{-7}$, $BR(B \rightarrow a_2 e^+ e^-)=(2.35^{+ 0.36}_{-0.27}{}^{+0.43}_{-0.38})\times 10^{-8}$, $BR(B \rightarrow a_2 \mu^+ \mu^-)=(1.28^{+ 0.21}_{-0.15}{}^{+0.23}_{-0.21})\times 10^{-8}$, $BR(B \rightarrow f_2 e^+ e^-)=(1.23^{+ 0.18}_{-0.14}{}^{+0.22}_{-0.20})\times 10^{-8}$, and $BR(B \rightarrow f_2 \mu^+ \mu^-)=(0.65^{+ 0.10}_{-0.07}{}^{+0.12}_{-0.11})\times 10^{-8}$. Our results on branching ratios are close to the that of the Ref.~\cite{Aliev:2019ojc} and Ref.~\cite{Zuo:2021kui} where $B\rightarrow T(2^+)ll$ are studied.

\section{Summary and conclusion}\label{IV} 

The present work provides the analyses for the $B_{(s)}\rightarrow T$ decays, where $T$ represents tensor mesons $K_2,~\phi_2,~f_2,~a_2$ mesons with $J^P=2^{-}$ quantum numbers. For these transitions, the relevant form factors were calculated for each decay within the light-cone QCD sum rules framework using the $B$-meson distribution amplitudes up to twist-four. The results for the form factors were then used to obtain the corresponding decay widths. Our calculations lead that all branching ratios for FCNC $B_{(s)} \rightarrow T l^+ l^-$ decays are at the order $10^{-7}\div 10^{-8}$, which can be obtained in future experiments. The branching ratio results attained in the present work for $B_{(s)}\rightarrow T(2^-) l^+l^-$ transitions are close to the results reported in Ref.~\cite{Aliev:2019ojc} and Ref.~\cite{Zuo:2021kui} for $B \rightarrow T(2^+)l^+l^-$ transitions. The results of our work may provide a testing ground for more precise possible future experiments.  

\appendix

\section{Distribution Amplitudes of the $B$-meson}
\label{app:LCDAs}

This section provides the LCDAs for the $B$ meson up to twist-four which are adopted in the present work. The $\bra{0} \bar{q}_{2}^{\alpha}(x) h_{v}^{\beta}(0) \ket{B_{q_2}(v)}$ matrix element in terms of DAs of the $B$ meson is determined as:
\begin{eqnarray}
\bra{0} \bar{q}_{2}^{\alpha}(x) h_{v}^{\beta}(0) \ket{B_{q_2}(v)}& =&
   -\frac{i f_B m_B}{4} \int^\infty_0 \text{d}\omega e^{-i\omega v\cdot x} \bigg\{
        (1 + \slashed{v}) \bigg[
            \phi_+(\omega) -g_+(\omega) \partial_\sigma \partial^\sigma
            +\left(\frac{\bar{\phi}(\omega)}{2}\right.
            \nonumber\\
            &-&\left. \frac{\bar{g}(\omega)}{2} \partial_\sigma \partial^\sigma\right) 
        \gamma^\mu \partial_\mu
        \bigg] \gamma_5
    \bigg\}^{\beta\alpha}.
    \label{eq:BLCDAs2pt}
\end{eqnarray}
In Eq.~(\ref{eq:BLCDAs2pt}), $v_{\mu}$ represents four velocity of the $B$-meson, $\partial_{\alpha}=\frac{\partial}{ \partial q^{\alpha}}$ and $q^{\alpha}=\omega v^{\alpha}$. The $\bar{\phi}(\omega)$ and $\bar{g}(\omega)$ are given as
\begin{eqnarray}
\label{eq:def:barred-LCDAs}
    \bar{\phi}(\omega) & =&\int_0^{\omega} \text{d}\xi\, \left(\phi_+(\xi) - \phi_-(\xi)\right)\,,\\
    \bar{g}(\omega) & =& \int_0^{\omega} \text{d}\xi\, \left(g_+(\xi) - g_-(\xi)\right)\,.
\end{eqnarray}
In our calculation we applied the model II A of Ref.~\citep{Braun:2017liq} for the two particle $B$-LCDAs $\psi_+$, $\psi_-$ and $g_+$ whose explicit expressions are given in Eqs.~(5.22) and (5.23) of Ref.~\citep{Braun:2017liq}.

Since there is no available model expression for $g_-$, we have used Wandzura-Wilczek (WW) approximation:
\begin{eqnarray}
\label{eq:gmWW}
 g_-^{WW}(\omega)
        & =&  +\frac{1}{4} \int_0^{\omega} \mathrm{d}\eta_2 \, \int_0^{\eta_2} \mathrm{d}\eta_1 \, \left[ \phi_+ (\eta_1) - \phi_-^{WW} (\eta_1) \right]  - \frac{1}{2} \int_0^{\omega} \mathrm{d}\eta_1 \, (\eta_1 - \bar{\Lambda}) \phi_-^{WW} (\eta_1)   \,,  \\ 
 \phi_-^{WW} (\omega)& =&  \int_{\omega}^{\infty} \mathrm{d}\eta_1 \, \frac{\phi_+ (\eta_1)}{\eta_1} \, ,
\end{eqnarray}
and
\begin{eqnarray}
 g_-^{WW}(\omega) =  \frac{\omega (3\lambda_B-\omega)^3 }{48 \lambda_B^3}\, \theta(3\lambda_B-\omega) \, ,
\label{eq:gmWWexplicit}
\end{eqnarray}
with $\theta(x)$ being the Heaviside step function~\cite{Gubernari:2018wyi}. $\lambda _E^2$, $\lambda _H^2$ and $\lambda _B$ present in the above equations are given in Sec.~\ref{III} as inputs.

\section{The normalization factor $A_{(T)}$ and coefficients $C_n^{(T,\psi_{\textrm{2p}})}$ appearing in LCSRs }
\label{app:coefficients}

After detailed calculations for the factors $A_{T}$ we find
\begin{eqnarray}
A_{A} &=&  \frac{f_T m_T^3}{m_B(m_B+m_T)}, \nonumber\\
A_{V_1} &=&   \frac{f_T m_T^3 (m_B+m_T)}{2 m_B}, \nonumber\\
A_{V_2 } &=&  - \frac{2 f_T m_T^3}{m_B (m_B+m_T)},\nonumber\\
A_{V_{30} } &=&   - \frac{{ f_T m_T^4}}{m_B} \, ,  \nonumber\\
A_{T_1} &=& - \frac{{f_T m_T^3}}{m_B} \, , \nonumber\\
A_{T_{23A}} &=& A_{(T_{23B} )}= \frac{{2 f_T m_T^3}}{m_B}  \, .
\end{eqnarray}
The (non-vanishing) coefficients in $C_n^{\psi}$ for the two particle DAs to the relevant invariant functions $\Pi_i$ are obtained as follows:

For the form factor $ A$: 
\begin{eqnarray}
C^{(A,\phi_+)}_1     & = &  \frac{x}{2}  \,,  \nonumber\\
C^{(A,\bar{\phi})}_2 & = & - \frac{m_{q_1} x}{2}  \,,\nonumber \\
C^{(A,g_+)}_2        & = &  4 x , \nonumber\\
C^{(A,g_+)}_3        & = &  -4 m_{q_1}^2 x \,, \nonumber\\
C^{(A,\bar{g})}_3    & = & - 4 m_{q_1} x   \, , \nonumber\\  
C^{(A,\bar{g})}_4    & = &  12 m_{q_1}^3 x \,.
\label{eq.B1}
\end{eqnarray}

For the form factor $V_{1}$:  
\begin{eqnarray}
C^{(V_{1},\phi_+)}_1     & = &  -\frac{x \left( q^2-(m_B\bar{x}-m_{q_1})^2 \right) }{4\bar{x}}  \,,\nonumber   \\
C^{(V_{1},\bar{\phi})}_1 & = & -\frac{x m_{q_1} }{4\bar{x}}   ,\nonumber \\
C^{(V_{1},\bar{\phi})}_2 & = &  \frac{x m_{q_1} \left( q^2-(-m_{q_1} + m_B\bar{x})^2 \right) }{4\bar{x}}  \,, \nonumber\\
C^{(V_{1},g_+)}_1        & = & \frac{2x}{\bar{x}}   ,  \nonumber\\
C^{(V_{1},g_+)}_2        & = & -\frac{2 x (q^2 - m_B \bar{x} (-m_{q_1} + m_B \bar{x}))}{\bar{x}}    \,, \\
C^{(V_{1},g_+)}_3        & = & \frac{2x m_{q_1}^2 \left( q^2-(-m_{q_1} + m_B\bar{x})^2 \right) }{\bar{x}}  \,,\nonumber \\
C^{(V_{1},\bar{g})}_2    & = & \frac{2x (-m_{q_1}+2 m_B \bar{x})}{\bar{x}}  \,,  \nonumber\\
C^{(V_{1},\bar{g})}_3    & = & \frac{2  m_{q_1}x (2m_{q_1}^2 + q^2 - m_B^2 \bar{x}^2)}{\bar{x}}  \,, \nonumber \\
C^{(V_{1},\bar{g})}_4    & = & -\frac{ 6 m_{q_1}^3 x (q^2- (-m_{q_1}+m_B\bar{x})^2 )}{\bar{x}}  \,.
\end{eqnarray}

For the form factor $V_{2}$:  
\begin{eqnarray}
C^{(V_{2},\phi_+)}_1     & = & x (1-2 \bar{x})   \,, \nonumber \\
C^{(V_{2},\bar{\phi})}_2 & = & -x (-m_{q_1}-2m_B\bar{x}+2m_B\bar{x}^2 )  \,,\nonumber\\
C^{(V_{2},g_+)}_2        & = & 8 x  (1-2 \bar{x}) \,,\nonumber\\
C^{(V_{2},g_+)}_3        & = & -8m_{q_1}^2 x  (1-2 \bar{x})  \,,\nonumber\\
C^{(V_{2},\bar{g})}_3    & = & - 8 x (-m_{q_1}-4m_Bx\bar{x} ) \,,\nonumber\\
C^{(V_{2},\bar{g})}_4    & = & 24 m_{q_1}^2 x (-m_{q_1}-2m_Bx\bar{x} ) \,.
\end{eqnarray}

For the form factor $V_{30}$: 
\begin{eqnarray}
C^{(V_{30},\phi_+)}_1     & = & \frac{x q^2 (2 {\bar x}-3)}{4}  \,, \nonumber  \\
C^{(V_{30},\bar{\phi})}_2 & = &  \frac{ x q^2 (2 x (1+x)m_B + m_{q_1}) }{4}  \,, \nonumber  \\
C^{(V_{30},g_+)}_2        & = & -2 q^2 x (1+2 x )   \,,  \nonumber  \\
C^{(V_{30},g_+)}_3        & = &  2 q^2 m_{q_1}^2 x  (1+2 x ) \,, \nonumber  \\
C^{(V_{30},\bar{g})}_3    & = & {2 x q^2 (  4 m_B x (1+x) + m_{q_1})} \,, \nonumber  \\
C^{(V_{30},\bar{g})}_4    & = & {6 m_{q_1}^2 q^2 x (-m_{q_1} -2m_B x (1+x))}  \,.
\end{eqnarray}

For the form factor $T_{1}$: 
\begin{eqnarray}
C^{(T_{1},\phi_+)}_1     & = & x (m_B \bar{x} - m_{q_1})/2    \,, \nonumber \\
C^{(T_{1},\bar{\phi})}_2 & = & - x m_{q_1}(m_B \bar{x}-m_{q_1})/2  \,,\nonumber \\
C^{(T_{1},g_+)}_2        & = & 2 x (2 m_B \bar{x} - m_{q_1})    \,, \nonumber \\
C^{(T_{1},g_+)}_3        & = & -4 x m_{q_1}^2 (m_B \bar{x} - m_{q_1})     \,,\nonumber \\
C^{(T_{1},\bar{g})}_2    & = & 4 x  \,, \nonumber \\
C^{(T_{1},\bar{g})}_3    & = & - 4 x \bar{x}m_B m_{q_1}     \,,\nonumber \\
C^{(T_{1},\bar{g})}_4    & = &  12 x m_{q_1}^3 (m_B \bar{x}-m_{q_1})       \,.
\end{eqnarray}

For the form factor $T_{23A}$:
\begin{eqnarray}
C^{(T_{23A},\phi_+)}_1     & = & x (m_B \bar{x} - m_{q_1})   \,, \nonumber \\
C^{(T_{23A},\bar{\phi})}_2 & = &  x (-m_{q_1}(m_B \bar{x} - m_{q_1})-2 x q^2)     \,,\nonumber \\
C^{(T_{23A},g_+)}_2        & = & 4 x (-m_{q_1}+2m_B \bar{x})      \,, \nonumber \\
C^{(T_{23A},g_+)}_3        & = & -8 m_{q_1}^2 x ( - m_{q_1} + m_B \bar{x})          \,, \nonumber \\
C^{(T_{23A},\bar{g})}_2    & = & 8 x     \,,\nonumber \\
C^{(T_{23A},\bar{g})}_3    & = &   8 x (-\bar{x} m_B m_{q_1} -4 x q^2)    \,,\nonumber \\
C^{(T_{23A},\bar{g})}_4    & = & -24 x m_{q_1}^2 (-\bar{x} m_B m_{q_1} +m_{q_1}^2-2 x q^2)     \,. 
\end{eqnarray}

For the form factor $T_{23B}$:
\begin{eqnarray}
C^{(T_{23B},\phi_+)}_1     & = & -x (x m_B + m_{q_1})      \,,\nonumber \\
C^{(T_{23B},\bar{\phi})}_1 & = &  -\frac{x^2}{\bar{x}}    \,, \nonumber \\
C^{(T_{23B},\bar{\phi})}_2 & = & \frac{x\hat{x}_1}{\bar{x}}          \,,\nonumber \\
C^{(T_{23B},g_+)}_2        & = & 4 x(-2 x m_B - m_{q_1})    \,, \nonumber \\
C^{(T_{23B},g_+)}_3        & = & 8 x m_{q_1}^2 (x m_B + m_{q_1})  \,, \nonumber \\
C^{(T_{23B},\bar{g})}_2    & = &   \frac{8 x (1 -3 x)}{\bar{x}}       \,, \nonumber \\
C^{(T_{23B},\bar{g})}_3    & = &   \frac{8 x^2 \hat{x}_2 }{\bar{x}}   \,,\nonumber \\
C^{(T_{23B},\bar{g})}_4    & = & -\frac{24 m_{q_1}^2 x \hat{x}_1}{\bar{x}}       \,.
\label{eq.B7}
\end{eqnarray}
In these results, $\hat{x}_1=m_{q_1}^2(1-2x) + m_B m_{q_1} x \bar{x}+x(m_B^2\bar{x}^2+q^2(2x-1))$ and $\hat{x}_2=2\bar{x}^2 m_B^2+\bar{x}m_B m_{q_1}-2q^2+4 x q^2+m_{q_1}^2$. The  Eqs.~(\ref{eq:T2}) and (\ref{eq:T3}) define $T_{23A}$ and $T_{23B}$.

%\section*{ACKNOWLEDGEMENTS}

%%%%%%%%%%%%%%%%%%%%%%%%%%%%%%
%\vspace{2mm} {\it Acknowledgments.}--- {\small .}

%%%%%%%%%%%%%%%%%%%%%%%%%%%%%%%%%%%%%%%%%%%%%%%%%%%%%%%%%%%%%%%%%%%%%%%%%%%%%%%%%%

\end{document}